# Tunable huge asymmetric transmission in mid-infrared range through a monolayer graphene chiral metasurface

Qi Wang[1] , Haotian Hao[2] , Yuyuan Zhang[3] , Wei Xin[*]

**Abstract:**

Metasurface has witnessed an urgent need for reconfigurable chiral manipulation recently, while the graphene-based devices attracted significant attention from researchers due to their inherent tunable properties. Here, an L-shaped metasurface composed of three hollow monolayer graphene resonant rings is proposed. A huge dual-frequency asymmetric transmission (AT) in the mid-infrared band (7800 nm-9000 nm) with transmission efficiency peaks of 0.14 and 0.26 is revealed, which is optimal of its similar monolayer counterparts in the current band. By changing the Fermi level of graphene and the relaxation time of its carriers, the AT spectrum including the positions and widths of the peaks can be effectively modulated. More importantly, the structure designed here gives priority to the machinability of the metasurface configuration, ensuring its practicality and promising applications in chiral applications in the future.

## 1. Introduction

In the past decade, the chiral metamaterials have shown great potential in the manipulation of electromagnetic waves, such as in circular dichroism [1-5], optical activity [6-8], and asymmetric transmission (AT) [9-11], which has attracted huge attention from researchers in applications. Among these phenomena, the planar chiral metamaterials with AT characteristics are attractive especially in chiral detectors, circular polarization converters, linear polarization rotators, and so on[12-14]. Although, metamaterials based on metal materials and dielectric materials have achieved significant AT [15, 16], the chiral effects of these metasurfaces are fixed once the structures have been fabricated[17].

In order to achieve dynamic chiral effects, some active materials, including phase change materials such as Ge2Sb2Te5 (GST) [18, 19]and VO2[20, 21], two-dimensional materials like black phosphorus [22] and graphene [23, 24], as well as liquid crystal [25], have been employed in traditional static metasurfaces. Among them, as a 2D material, graphene has excellent properties such as ultra-high electron mobility, tunable Fermi energy, low resistivity, good mechanical flexibility and stability and so on [26, 27] , which is a promising alternative in the preparation of AT metamaterials. Moreover, it presents an inherent advantage in device miniaturization due to its limited size. However, limited by the enantio-sensitive plasma excitation intensity, the reported AT of monolayer graphene metasurfaces is usually low and it needs to be improved especially in the infrared region due to large application requirements [28-30].

In this paper, an L-shaped graphene chiral metasurface composed of three hollow resonant rings is proposed, and its AT characteristics are studied by numerical simulation in the mid-infrared band. Two AT peaks with 6.03% and 11.76% efficiencies can be obtained at 8125 nm and 8360 nm respectively. Furthermore, by adjusting the Fermi energy of graphene, the dynamic regulation of the

AT is realized. In this process, we believe the asymmetric excitation electric field induced by the enantiomers-sensitive plasmon polaritons is dominant. Our work fully considers the convenient machinability of the graphene metasurface configuration, which provides theoretical support for the practical applications of polarization manipulation, biosensors and chiral detection in the infrared field. The relevant phenomena can be verified and carried out in the future experiments.

## 2. Model and theoretical framework

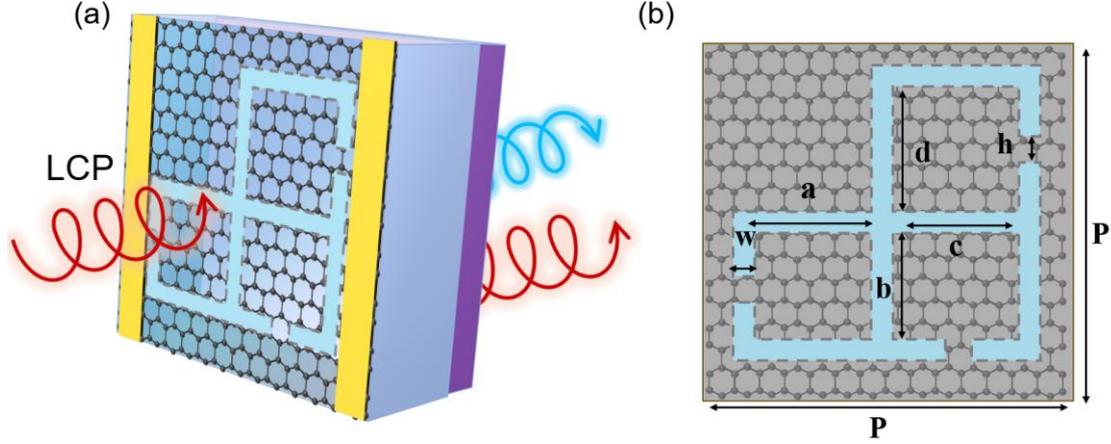

**Fig. 1** (a) An artistic rendering of the asymmetric transmission in the graphene planar chiral metasurface. (b) A unit cell with structure parameters. The width of all the hollows is w=15 nm. The length of the gaps is h=10 nm. The length of hollow are a=100 nm, b=80 nm, c=70 nm, and d=90nm.

Figure 1 is a schematic diagram of a graphene chiral metasurface, which is composed of a single-layer hollow graphene and a dielectric substrate. The hollow part is a L-shaped structure composed of three open resonant rings that overlap vertically. The left-handed circularly polarized light and the right-handed circularly polarized light in the mid-infrared band are incident from the positive and negative sides of the graphene metasurface, respectively, as shown in Fig. 1 (a). Figure 1 (b) shows a cell of a graphene chiral metasurface with geometric parameters. The grid is graphene, and the blank area is the excavated pattern. The period of the structure is P = 300 nm, which ensures that the structure will not be diffracted in the spectral range studied under normal incident conditions.

In this study, the asymmetric transmission of circularly polarized light in the 8000 nm-9000 nm band is studied by finite element analysis. The Fermi energy of graphene is between 0.84 eV and 0.94 eV. The periodic boundary condition is applied at the boundary of the graphene cell, and the graphene is set as the transition boundary condition. The conductivity of graphene can be obtained using the Kobe equation given below[31].

$$\sigma = \frac{ie^2 E_f}{\pi\hbar(\omega+i\tau^{-1})} + \frac{ie^2}{4\pi\hbar}\ln\left[\frac{2E_f - \hbar(\omega+i\tau^{-1})}{2E_f + \hbar(\omega+i\tau^{-1})}\right] + \frac{ie^2 k_B T}{\pi\hbar^2(\omega+i\tau^{-1})}\ln\left[\exp\left(-\frac{E_f}{k_B T}\right)+1\right] \qquad (1)$$

In the equation, $\omega$ is the frequency of the incident electromagnetic wave, $\tau$ is the relaxation time of the carriers, $\mu$ is the carrier mobility, $E_f$ is the graphene's Fermi energy, $T$ is the ambient temperature, $e$ is the charge of the electron, $\hbar$ is the reduced Planck constant, and $k_B$ is the Boltzmann constant.

When the circularly polarized light is incident into the object along the positive direction of the z-axis, the electric field can be expressed as follow:

$$\vec{E_i}(\vec{r},t) = \begin{pmatrix} I_{LCP} \\ I_{RCP} \end{pmatrix} e^{i(kz-\omega t)} \quad (2)$$

Where, I represents the incident light, $\omega$ is the incident light frequency, k is the incident light wave vector. Moreover $I_{LCP}$ and $I_{RCP}$ are the complex amplitudes of the incident left-handed circularly polarized (LCP) light and right-handed circularly polarized (RCP) light, respectively. Similarly, the electric field of the emitted light can be represented in the following equation:

$$\vec{E_t}(\vec{r},t) = \begin{pmatrix} T_{LCP} \\ T_{RCP} \end{pmatrix} e^{i(kz-\omega t)} \quad (3)$$

Here, $T$ represents the emitted light. By using the Jones matrix to establish a connection between the incident light field and the transmitted light field, the transmission matrix of the electromagnetic wave propagating along the positive direction of the z-axis can be obtained in the following fashion [32].

$$\begin{pmatrix} T_{LCP} \\ T_{RCP} \end{pmatrix} = \begin{pmatrix} t_{LL} & t_{LR} \\ t_{RL} & t_{RR} \end{pmatrix} \begin{pmatrix} I_{LCP} \\ I_{RCP} \end{pmatrix} \quad (4)$$

$$T_{cric}^{f} = \begin{pmatrix} t_{LL} & t_{LR} \\ t_{RL} & t_{RR} \end{pmatrix} \quad (5)$$

In the superscript, $f$ represents the incident direction along the positive z-axis direction. In the subscript, "circ" represents the circularly polarized light as the base vector. The graphene metasurface designed in this study does not involve magneto-optical materials, and light propagation on the metasurface follows the reciprocity theorem. Therefore, when the light wave is incident along the negative direction of the z-axis, the following equation denotes the corresponding transmission matrix:

$$T_{cric}^{b} = \begin{pmatrix} t_{LL} & t_{RL} \\ t_{LR} & t_{RR} \end{pmatrix} \quad (6)$$

Superscript $b$ denotes the negative direction of the incident direction along the z-axis. Therefore, the transmitted light intensity is given in the following equations when the right-handed circularly polarized light is incident on the metasurface from both positive and negative directions:

$$T_{RCP}^{f} = |t_{LR}|^2 + |t_{RR}|^2 \quad (7)$$

$$T_{RCP}^{b} = |t_{RL}|^2 + |t_{RR}|^2 \quad (8)$$

Asymmetric transmission refers to the phenomenon that the transmittance differs when the electromagnetic wave is in the opposite direction. For the circular base, it can be expressed as:

$$\Delta T = \Delta_{cric}^{LCP} = |t_{LR}|^2 - |t_{RL}|^2 = -\Delta_{cric}^{RCP} \quad (9)$$

When ΔT has a nonzero, the device has asymmetric transmission properties.

## 3. Results

Numerical calculations with the commercial software COMSOL can simulate and calculate the changes in the graphene metasurface transport matrix. The matrix coefficient is the square of the absolute value of the corresponding coefficient of the Jones matrix, $T_{ij} = |t_{ij}|^2$. Figure 2 (a) shows the calculation results of four transmission matrix elements when the Fermi energy of graphene super surface is 0.9eV. The results show that the direct transmittance $t_{LL}$ and $t_{RR}$ of the left-handed circularly polarized light and the right-handed circularly polarized light are always equal in the wavelength range of 7800 nm-9000 nm, and do not change with the wavelength, which proves that the designed graphene super surface does not have the circular dichroism of three-dimensional chiral characteristics. The polarization conversion rates $T_{RL}$ and $T_{LR}$, show significant differences, indicating that the metasurface had better asymmetric transmission performance. According to Eq. (9), the asymmetric transmission curve of the right circularly polarized light was calculated in this paper, as shown in Fig. 2(b), The metasurface has two asymmetric transmission peaks at 8208 nm and 8688 nm, with 0.14 and 0.26 transmission, respectively.

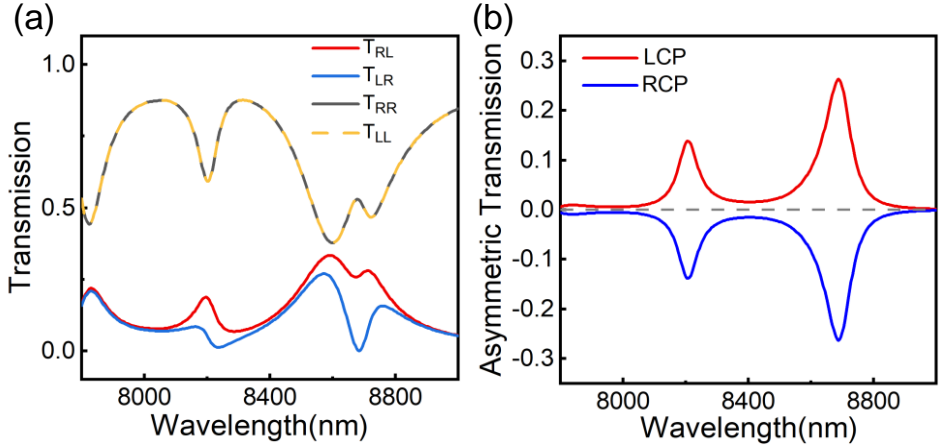

**Fig. 2** (a) Calculated squared moduli $T_{ij} = |t_{ij}|^2$ of the graphene planar chiral metasurface for forward propagation. (b) Asymmetric transmission for right-handed circularly polarized waves propagating along forward direction. The Fermi energy is fixed at 0.90 eV.

The physics of asymmetric transport phenomena is explained by analyzing the electromagnetic field coupling mechanisms in metamaterials. The delay wave generated by strong electromagnetic coupling in the metamaterial contains a component perpendicular to the polarization direction of the incident wave. This perpendicular component causes the entire vibrational surface of the transmitted wave to rotate, resulting in polarization conversion. Electromagnetic waves on this metasurface form a current loop, creating magnetic dipoles that excite an induced magnetic field. When the induced magnetic field is parallel to the direction of the electric field of the incident electromagnetic wave, the two will cross-couple, resulting in a change in the polarization direction of the incident electromagnetic wave, that is, from left circularly polarized light to right circularly polarized light, resulting in asymmetric transmission phenomenon. Furthermore, when the induced magnetic field is perpendicular to the direction of the incident electric field, there is no cross-coupling between them and, therefore, no deflection of the polarization direction of the electromagnetic wave. Figure 3 shows the induced electric field distribution of graphene layer when LCP and RCP waves pass through the graphene chiral metasurfaces in the forward direction at 8208 nm and 88688 nm, respectively. And the induced electric vector field is indicated by red arrows. The polarization conversion intensity is determined by the electric field strength in the y direction, when the incident light is LCP, the electric field in the positive direction of the y-axis promotes the polarization conversion, the electric field in the negative direction of the y-axis suppresses the polarization conversion, and the reverse is true for RCP. At 8208 nm, the electric field generated along the direction of the promoted polarization transition when the incident light is LCP is smaller than when RCP light is incident, so the $T_{LR}$ is large and the $T_{RL}$ is small. Likewise, at a wavelength of 8688 nm, the electric field along the direction of promoting polarization conversion generated when the incident light is LCP is significantly smaller than the electric field generated when RCP is incident light, thus producing an obvious asymmetric transmission peak.

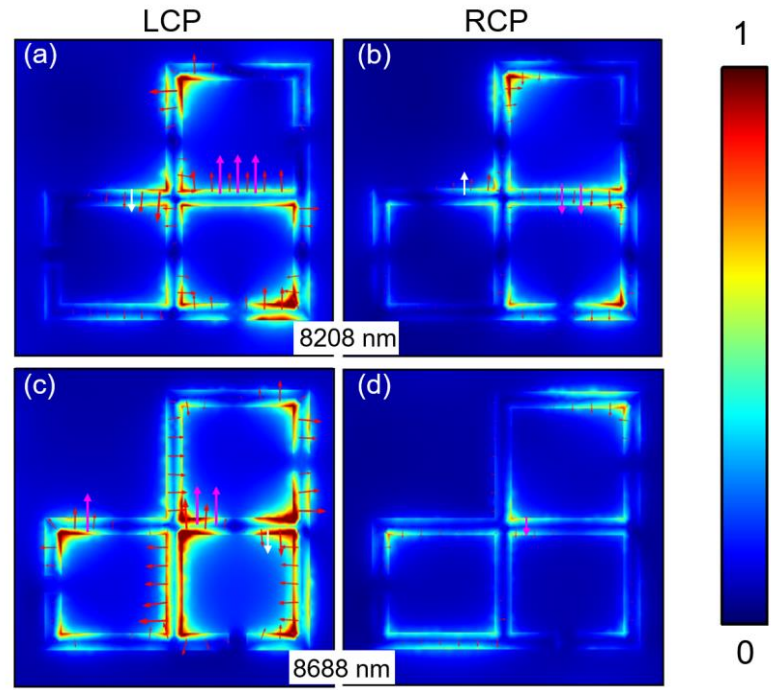

**Fig. 3** Maps of the magnitude of the resonance-induced current density and the local electric distribution in the hole area of the meta-surface (a), (c) left-handed polarized and (b), (d) right-handed polarized waves propagating along the forward direction at the wavelengths of 8208 nm and 8688 nm, respectively. The Fermi energy is fixed at 0.90 eV.

In order to show the dynamic modulation ability of graphene metasurfaces for asymmetric transmission, we analyzed the influence of graphene's Fermi energy $E_f$ change on the metasurfaces asymmetric transmission value, as shown in Fig. 4(a). With the decrease of graphene's Fermi energy, the asymmetric transmission spectral line has a redshift, while the shape and amplitude of the spectral line remain basically unchanged. This means that by changing the Fermi energy of graphene, it can maintain the asymmetric transmission performance of the metasurface and make it work in different band.

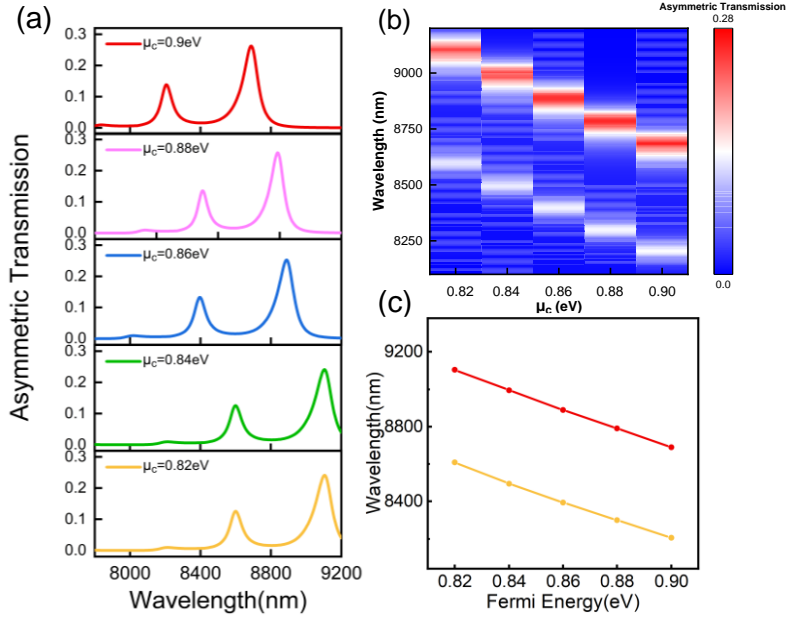

**Fig. 4** (a) The asymmetric transmission for forward propagation at different Fermi energies. (b) (c) Asymmetric transmission as a function of wavelength and Fermi energy for forward propagation.

In order to further explore the relationship between the Fermi energy of graphene and the asymmetric transmission effect, we explored the change of wavelength at the two asymmetric transmission peaks with the Fermi energy, as shown in Figs. 4(b) and 4(c). The results show that when the Fermi energy changes from 0.82 eV to 0.90 eV, the peak position undergoes blue-shift. This phenomenon primarily occurs due to the variation in the wavelength of the surface plasmon formed by the incident light and graphene metasurface resonance at different Fermi energys. The resonant wavelength $\lambda$ has the following relationship with Fermi energy $E_f$:

$$\lambda = \frac{2\pi c}{\omega} \propto \sqrt{\frac{2\pi^2 \hbar L_g}{\alpha_0 c E_f}} \qquad (10)$$

Here, $\alpha_0 = \frac{e^2}{\hbar c}$ is the fine structure constant, $L_g$ represents the resonant characteristic length of the graphene hollow pattern. Therefore, Eq. (10) clearly shows that the increase of Fermi energy $E_f$ will cause to the blue shift of these resonant wavelength.

The linear dependence of the resonance wavelength on the Fermi energy is attributed to the relation between the optical response and the Fermi energy of the graphene. The real part of the graphene's permittivity has a linear relationship with changing Fermi energy, while its imaginary part almost keeps constant in a large wavelength range, as shown in Figs. 5(a) and (b).

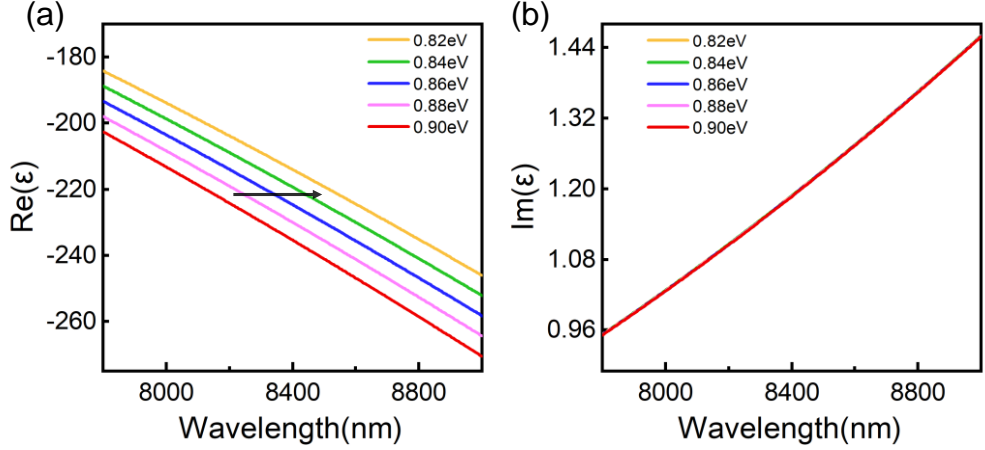

**Fig. 5** (a) The real part and (b) the imaginary part of the graphene's permittivity as a function of the Fermi energy.

The carrier mobilities of graphene samples prepared by different processes are different. In general, the carrier mobility of graphene is between 3000 cm$^2$/Vs and 10000 cm$^2$/Vs. The differences in carrier mobilities lead to the change in relaxation times, which affect the permittivities of graphene. Therefore, this study explored the change of asymmetric transmission with relaxation time at the Fermi energy of 0.90 eV, as shown in Fig. 6. With the decrease of relaxation time, the peak amplitude of the asymmetric transmission decreases.

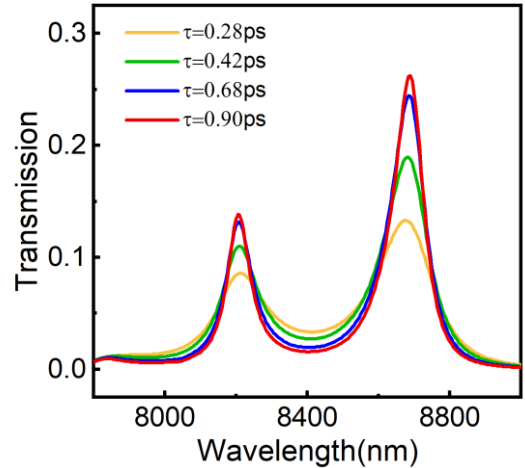

**Fig. 6** Asymmetric transmission spectra for different intrinsic relaxation times, where the Fermi energy is fixed at 0.9 eV.

**4. Conclusions**

In this work, an L-shaped monolayer graphene metasurface consisting of three hollow resonant rings was designed. The metasurface showed strong AT characteristics for circular polarized light in the 7800-9000 nm band, with the peak transmission values reaching 0.14 at 8208 nm and 0.26 at 8688 nm, respectively. Through the analysis of the electric field intensity and current distribution, the chiral excitation of plasmon is found the root cause of the phenomenon. By adjusting the Fermi energy and relaxation time of graphene, the AT performance can be tuned with reconfigurable characteristics. Moreover, the device structure designed here is simple enough for its actual processing, which presents great significance in applications in the future.